\documentclass[12pt]{article} % single column, double spaced
\usepackage{ol2}
\usepackage[draft]{hyperref}
\usepackage{amsmath}
\begin{document}
%\twocolumn[ %% activate for two-column option
\title{A lossless metamaterial with tunable permittivity and its application as a compact phase shifter}
\author{Yong Zeng, Qi Wu and Douglas H. Werner}
\address{Department of Electrical Engineering, Pennsylvania
State University, University Park, PA 16802}
\begin{abstract}
In this Letter, we propose a new type of lossless metamaterial
whose effective permittivity is tunable from negative to positive
values. Its optical response is studied analytically and
numerically. We further demonstrate that this tunable metamaterial
can significantly modulate the phase of an incident pulse with
negligible reflection loss, functioning as an efficient phase
shifter.
\end{abstract} \ocis{160.3918, 230.0230, 050.5080} %]

\newpage

Metamaterials are man-made composites engineered on a
subwavelength scale, designed to produce an optimized combination
of electromagnetic properties that may not be readily available in
nature \cite{marques,solymar,zhang}. They have applications
ranging from perfect imaging lenses \cite{john} to electromagnetic
invisibility \cite{pendry,leonhardt,schurig}. One rapidly emerging
branch of this field concerns tunable hybrid metamaterials in
which plasmonic composites are integrated with natural materials
whose electrical, optical, thermal or mechanical properties can be
externally controlled
\cite{padilla,chen3,khoo,werner,wang,kwon,bossard,chen1,xiao,yoshida,nemec,driscoll,liu}.
The resultant media have tunable effective material parameters,
and their electromagnetic response can be adjusted in real time.
This dynamic tuning ability makes metamaterial devices more
flexible and versatile. For instance, frequency-agile terahertz
metamaterials have been fabricated recently by incorporating
silicon in critical regions of metallic split-ring resonators.
Through photoexcitation of free carriers in the semiconductor, the
resonant frequency of the hybrid metamaterial can be tuned
considerably under specific excitations \cite{chen1}. It was
further demonstrated experimentally that this frequency-agile
metamaterial can modulate the phase or amplitude of the incident
electromagnetic pulse, and therefore can function as a phase
modulator \cite{chen2}. Unfortunately, owing to the intrinsic
ohmic loss as well as the resonant nature of plasmonic
nanostructures, these tunable metamaterials suffer significant
electromagnetic absorption and thus degraded performance. To date
this has been a serious drawback preventing the implementation of
tunable metamaterials into practical devices.

One approach to alleviate or even eliminate the metal absorption
loss is to introduce gain media into the metamaterial composites
\cite{bergman,Oulton}. This method has been applied to achieve
spasers \cite{zheludev,noginov} and lossless negative permittivity
metamaterials \cite{fu,Bratkovsky,zeng}. In this Letter, we will
integrate the hybrid metamaterials with gain media to realize
lossless metamaterials with tunable permittivities. The
metamaterial considered here consists of coated nanospheres with a
gain core dispersed in a homogeneous host material whose
permittivity can be controlled externally. Because the
characteristic size of the nanoparticle is far smaller than the
free-space wavelength, we study its optical response using the
electrostatic (long-wavelength) approximation. We demonstrate that
the hybrid metamaterial is always lossless at the wavelength where
the absorption cross section of the constitutive particle
vanishes. Furthermore, modifying the permittivity of the host
medium will tune the real part of the effective permittivity of
the composite. Under certain conditions, the resultant bulk
permittivity is tunable from negative, through zero, to positive
values. The analytical predictions are then validated by comparing
them with those of rigorous full-wave simulations. Finally, we
propose that a thin slab of the tunable metamaterial can linearly
manipulate the phase of an excitation pulse with negligible
reflection, functioning as a compact phase shifter
\cite{chen2,he}.

\begin{figure}[t]
\centering
\includegraphics[width=0.7\textwidth]{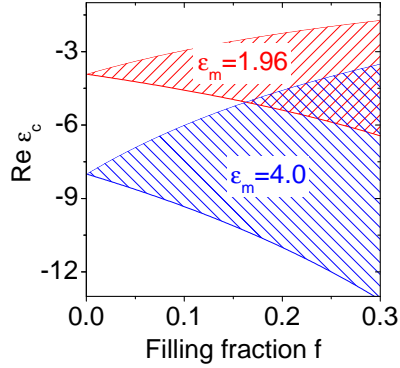}\vspace*{-9.0cm}
\caption{A graphical display of Eq. (\ref{eq3}) with
$\epsilon_{m}$ in the region $[1.96, 4.0]$. $\epsilon_{b}$ is
negative in each of the shaded regions for a specific
$\epsilon_{m}$, and $\epsilon_{b}$ is negative in the overlapped
triangle, regardless of $\epsilon_{m}$.} \label{fig1}
\end{figure}

We start by considering a coated sphere with inner radius $r_{1}$
and outer radius $r_{2}$. When its size is much smaller than the
free-space wavelength, an effective permittivity can be employed
to describe its classical optical response \cite{bohren}
\begin{equation}
\epsilon_{c}=\epsilon_{2}\frac{\epsilon_{1}(1+2\rho)+2\epsilon_{2}(1-\rho)}{\epsilon_{1}(1-\rho)+\epsilon_{2}(2+\rho)},
\label{eq1}
\end{equation}
where $\rho=r_{1}^{3}/r_{2}^{3}$ is the volume fraction,
$\epsilon_{1}$ and $\epsilon_{2}$ are the permittivity of the core
and shell, respectively. Moreover, a composite consisting of these
particles has an effective bulk permittivity governed by the
Clausius-Mossotti relation \cite{bohren}
\begin{equation}
\epsilon_{b}=\epsilon_{m}\frac{\epsilon_{c}(1+2f)+2\epsilon_{m}(1-f)}{\epsilon_{c}(1-f)+\epsilon_{m}(2+f)},
\label{eq2}
\end{equation}
where $f$ is the filling fraction of the sphere, and
$\epsilon_{m}$ is the permittivity of the host medium which is
assumed to be positive and real-valued. This equation immediately
suggests that the imaginary part of $\epsilon_{c}$ must vanish to
result in a lossless $\epsilon_{b}$ since $\epsilon_{m}$ is real
valued \cite{zeng}. Furthermore, requiring $\epsilon_{b}\leq 0$
leads to the condition
\begin{equation}
\epsilon_{m}\frac{2+f}{f-1}<\mathrm{Re}\:\epsilon_{c}\leq\epsilon_{m}\frac{2f-2}{1+2f}.
\label{eq3}
\end{equation}
We graphically plot this inequality in Fig. \ref{fig1}, and assume
$\epsilon_{m}$ has a minimum of 1.96 and a maximum of 4.0. The
overlapped triangular region indicates the fully negative zone in
which a set of point $(f,\mathrm{Re}\:\epsilon_{c})$ always
results in a negative value of $\epsilon_{b}$. The vertex of the
triangle therefore corresponds to a critical filling fraction
$f_{c}$, and a smaller filling fraction may lead to a positive
value of $\epsilon_{b}$. In addition, the partial derivative of
$\epsilon_{b}$ with respect to $\epsilon_{m}$ can be derived from
Eq. (\ref{eq2})
\begin{equation}
\frac{d\epsilon_{b}}{d\epsilon_{m}}
=(1-f)\frac{(\epsilon_{c}+2\epsilon_{m})^{2}+2f(\epsilon_{c}-\epsilon_{m})^{2}}{\left[\epsilon_{c}(1-f)+\epsilon_{m}(2+f)\right]^{2}},
\label{eq4}
\end{equation}
which is always positive for any $f< 1$, implying that
$\epsilon_{b}$ is a monotonically increasing function of
$\epsilon_{m}$. This is not surprising because the optical
properties of the composite medium directly relate to the volume
average of the guest and host media in the electrostatic
approximation.

The observations above suggest that a design procedure consisting
of two nearly independent steps can be employed. In the initial
step, the first-order surface mode of a core-shell nanosphere is
nearly excited at the targeted frequency such that the resulting
$\epsilon_{c}$ has a vanishing imaginary part with a considerable
negative real part \cite{zeng}. An externally controllable host
medium is then introduced at the second step to tune the effective
bulk permittivity of the composite. To demonstrate this
easy-to-use design procedure, we consider the following realistic
example. The nanoparticles are silver coated spherical
semiconductor quantum dots (with optical gain) which are almost
identical to those studied in Ref.\cite{zeng} except for the
geometrical parameters: the radius of the metallic shell and the
semiconductor core is 1.38 nm and 8.62 nm, respectively. This
coated sphere is found to possess a lossless negative
$\epsilon_{c}$ at a wavelength of 834 nm. The host material in
this case is chosen to be a planar aligned nematic liquid crystal,
which possesses a large electro-optics response and has been
employed to achieve reconfigurable metamaterials with a
negative-zero-positive index of refraction in the optical regime
\cite{khoo,wang,kwon,bossard}. It should be emphasized that this
hybrid composite may be fabricated by a sputter doping approach
\cite{yoshida}. For linearly polarized light incident as an
extraordinary wave onto the liquid crystal host, its permittivity
depends on the director axis orientation angle $\phi$ with respect
to the incident wave vector \cite{khoo}
\begin{equation}
\epsilon_{m}=\frac{\epsilon_{\parallel}\epsilon_{\perp}}{\epsilon_{\parallel}\cos^{2}\phi+\epsilon_{\perp}\sin^{2}\phi},
\label{eq5}
\end{equation}
where $\epsilon_{\parallel}\approx 4.0$ ($\epsilon_{\perp}\approx
1.96$) is the permittivity for light polarized parallel
(perpendicular) to the director axis. Through modulating $\phi$
either electrically or optically, we can change $\epsilon_{m}$
from $\epsilon_{\perp}$ to $\epsilon_{\parallel}$ \cite{khoo}.

\begin{figure}[t]
\centering
\includegraphics[width=0.5\textwidth]{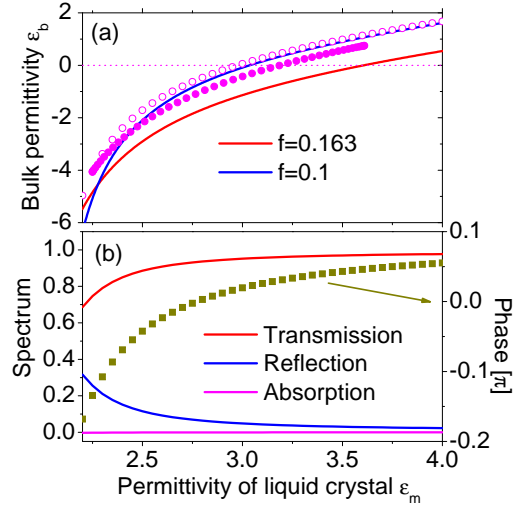}\vspace*{-3.5cm}
\caption{(a) The effective permittivity $\epsilon_{b}$ of the bulk
composite for two filling fractions $f$. Solid curves show
analytical results, while the circles represent full-wave
simulations. The empty circles correspond to an isotropic model of
the liquid crystal, while the filled circles correspond to an
anisotropic model. (b) The spectra of a thin metamaterial slab,
with a thickness of 34.7 nm, under normal incidence.} \label{fig2}
\end{figure}

Using Eq. (\ref{eq2}), the bulk permittivities $\epsilon_{b}$,
corresponding to two different filling fractions, are computed and
the results are plotted in Fig. 2(a) as solid curves. The
effective permittivities $\epsilon_{b}$ have vanishing imaginary
parts while the real parts can be monotonically tuned from
negative, through zero, to positive values with the increment of
$\epsilon_{m}$ of the liquid crystal. In addition, the zero
$\epsilon_{b}$ condition is given by
\begin{equation}
\epsilon_{m}=-\frac{\epsilon_{c}(1+2f)}{2(1-f)}. \label{eq6}
\end{equation}
Hence, a smaller filling fraction therefore requires a smaller
$\epsilon_{m}$. To validate the electrostatic results, we consider
the nanoparticles in a simple cubic lattice with lattice spacing
of 34.7 nm (corresponding to $f=0.1$). A full-wave finite-element
method is then used to simulate a single layer of lattice under
normal incidence \cite{comsol}. The calculated spectra are shown
in Fig. 2(b) and the absorption is always zero, consistent with a
lossless $\epsilon_{b}$. The transmission/reflection method is
finally applied to extract the effective permittivity
\cite{smith2}. The result is plotted in Fig. 2(a) with empty
circles, and is found to be in good agreement with its analytical
counterpart. A rigorous anisotropic treatment of the liquid
crystal elements, the same as that carried out in Ref.\cite{wang},
is also utilized. We avoid the complicated cross-polarization
coupling by carefully aligning the incident wave vector with the
liquid crystal director axis \cite{wang}. The corresponding
numerical result is plotted in Fig. 2(a) with filled circles, and
is nearly identical to its isotropic counterpart. Notice that the
value of $\epsilon_{m}$ of the anisotropic model is within $[2.25,
3.61]$, narrower than the isotropic model.

Our tunable permittivity metamaterials with compensated losses may
have a variety of applications, such as $\epsilon$-near-zero
materials \cite{silveirinha}. Of these, the phase shifter may be
the most interesting \cite{chen2,he}. To illustrate its
principles, we consider a thin slab, with thickness $d$ much
smaller than the free-space wavelength $\lambda$, embedded in a
vacuum. Under normal incidence its transmittance can be
approximated as
\begin{equation}
T\approx
1-\frac{d^{2}\pi^{2}}{\lambda^{2}}\left(\epsilon-1\right)^{2},
\label{eq7}
\end{equation}
with $\epsilon$ being the real-valued, negative or positive,
permittivity of the slab. Clearly the reflectance is negligible
for a modest $\epsilon$. Moreover, the phase of the transmission
is given by
\begin{equation}
\theta\approx \frac{d\pi}{\lambda}\left(\epsilon+1\right),
\label{eq8}
\end{equation}
implying that $\theta$ is a linear function of the permittivity:
It is negative when $\epsilon<-1$ and positive when $\epsilon>-1$.
Consequently, the transmitted light can be delayed or advanced.
The single layer slab studied above can serve as an example. As
shown in Fig. 2(b), $\epsilon_{b}=-1$ indeed results in a zero
$\theta$, and the accumulated phase difference
$\Delta\epsilon_{b}d\pi/\lambda$, obtained by increasing
$\epsilon_{m}$ from 2.0 to 4.0, is around $0.25\pi$. Notice that
the thickness of the slab is only 34.7 nm, indicating that it
operates as an efficient phase shifter. In addition, to achieve a
larger accumulated phase difference, we need to increase the slab
thickness which, however, decreases the transmittance. The linear
dependence of $\theta$ and quadratic dependence of transmittance
$T$ on $d$ suggest an approach to alleviate the reflection loss
considerably: Replace a single layer slab with multilayer slabs
while keeping the total thickness unchanged. It should be
mentioned that a similar method has been employed to improve the
imaging resolution of near-sighted superlenses
\cite{shamonina,podolskiy}.

In summary, we have described an innovative hybrid metamaterial,
formed by integrating gain medium with an externally controllable
host material. The composite exhibits a tunable effective
permittivity with compensated losses. It is further suggested that
a thin slab of this tunable material can function as an efficient
phase shifter whose transmittance is nearly 100\%.

This work was supported in part by the Penn State Materials
Research Science and Engineering Center under National Science
Foundation grant no. DMR 0213623.

\end{document}